\documentstyle [12pt]{article}
\newcommand{\be}{\begin{equation}}
\newcommand{\ee}{\end{equation}}
\newcommand{\ra}{\rightarrow}

\newcommand{\lsim}{\stackrel{<}{\sim}}
\begin{document}
\begin{titlepage}
\begin{flushright}
 IFUP-TH 38/93\\
  September 1993\\
  hep-th/9309034
\end{flushright}

\vspace{8mm}

\begin{center}

{\Large\bf  The algebraic structure of the

\vspace{4mm}

 generalized uncertainty principle }\\

\vspace{12mm}
{\large Michele Maggiore}

\vspace{3mm}

I.N.F.N. and Dipartimento di Fisica dell'Universit\`{a},\\
piazza Torricelli 2, I-56100 Pisa, Italy.\\
\end{center}

\vspace{4mm}

\begin{quote}
\hspace*{7mm} ABSTRACT. We show that a  deformation of the
{\mbox Heisenberg} algebra
which depends on a dimensionful parameter $\kappa$ is
the algebraic structure which underlies
the generalized uncertainty principle in quantum gravity.
The deformed algebra and therefore
the form of the
generalized uncertainty principle are fixed uniquely
by rather simple assumptions. The string theory result is reproduced
expanding our result at first order in $\Delta p/M_{\rm PL}$.
We also briefly comment on possible
 implications for Lorentz invariance at
the Planck scale.
\end{quote}
\end{titlepage}
\clearpage

In recent years it has been suggested that measurements in quantum
gravity are governed by
  a generalized uncertainty principle
\be\label{1}
\Delta x\ge \frac{\hbar}{\Delta p} +{\rm const.}\, G\Delta p\,
\ee
( $G$ is Newton's  constant). At energies much below the Planck mass
$M_{\rm PL}$ the extra term in eq.~(1) is irrelevant and the
Heisenberg relation is recovered. As we approach the Planck mass, this
term
becomes important and it is responsible for the existence of a minimal
observable length on the order of the Planck length.

The result (1) was first suggested in the context of string theory
[1-4], in the kinematical region where $2GE$ is smaller than the
string length (for a review, see~\cite{Cia}).
However, heuristic
arguments~\cite{MM1} suggest that this formula might
have a more general
validity in quantum gravity, and it is not necessarily related
to strings. It is therefore
natural to ask whether there is an algebraic
structure which reproduces eq.~(1) (or, more in general,
which reproduces the existence of a minimal observable length),
in the same way in which the
 Heisenberg uncertainty principle follows from the
 algebra $\left[ x,p\right] =i\hbar$.

In this Letter we answer (in the affirmative)
 this question. Our strategy is as follows.
Since it is relatively clear that no Lie algebra can reproduce eq.~(1),
we turn our attention to deformed algebras.
A deformed algebra is an
associative algebra where it is defined a commutator which is
{\em non-linear} in the elements of the algebra; and there is a
deformation parameter such that, in an appropriate limit, a
Lie algebra is recovered.

We therefore look for the most general deformed
algebra\footnote{We will not require the existence of a coproduct and
of an antipode, which would promote the deformed algebra to a quantum
algebra, see below.}
 which can be
constructed from coordinates $x_i$ and momenta $p_i\, (i=1,2,3)$.
We restrict the
range of possibilities making the following assumptions.
1) The three-dimensional rotation group is not deformed; the angular
momentum ${\bf J}$ satisfies the undeformed $SU(2)$ commutation
relations, and coordinate and momenta satisfy the undeformed
commutation relations $\left[ J_i,x_j\right] =i\epsilon_{ijk}x_k,
\left[ J_i,p_j\right] =i\epsilon_{ijk}p_k$.
2) The momenta commutes between themselves: $\left[ p_i,p_j\right]
=0$, so that also the translation group is not deformed.
3) The $\left[ x,x\right]$ and $\left[ x,p\right]$ commutators
depend on a deformation
parameter $\kappa$ with dimensions of mass. In the limit
$\kappa\ra\infty$ (that is, $\kappa$ much larger than any energy), the
canonical commutation relations are recovered.

Note that we are ready to accept a non-zero commutator between the
$x$'s. While this is at first surprising, we will see that
if $\kappa\sim$ Planck mass,  the non-commutativity shows up only
at the level of the Planck length. The idea is in the spirit of the
approach to the structure of spacetime at small distances suggested by
non-commutative geometry, pioneered
 by Connes~\cite{Con}, which
underlies much of the applications of quantum groups to
gravity.  One should also note that our formalism is not Lorentz
covariant. We will come back  to this important point below.

With these assumptions,
the most general form of the $\kappa$-deformed algebra is
\begin{eqnarray}
\left[ x_i ,x_j \right] &= & \frac{\hbar^2a(E)}{\kappa^2}
i\epsilon_{ijk}J_k\\
\left[ x_i , p_j \right]   &= & i\hbar\delta_{ij}f(E)\, .
\end{eqnarray}
Here $a(E)$ and $f(E)$ are real, dimensionless functions of
$E/\kappa$, and $E^2=p^2+m^2$;
 the angular momentum ${\bf J}$ is defined as  dimensionless, so
on the right-hand side the
dimensions are carried by $\hbar$ and $\kappa$ only.
The fact that this is the most general form compatible with our
assumptions is  clear from the following considerations:
the factors of $i$ are determined by the
condition of hermiticity of $x_i, p_i$ and $J_i$. The powers of
$\hbar$ and $\kappa$ are dictated by dimensional analysis. The
tensor $\epsilon_{ijk}$ in  eq.~(2)
appears because we assume that the
three-dimensional rotation group is undeformed and then it is the only
tensor antisymmetric in $i,j$; it must be
contracted with $J_k$ rather than $p_k$ or $x_k$
because of parity. One might also add  to the
right-hand side of eq.~(2) a term proportional to $x_ip_j-x_jp_i$.
(Note that $L_k=\epsilon_{ijk}(x_ip_j-x_jp_i)$
cannot be identified with the angular
momentum in the spinless case, since it does not satisfy the $SU(2)$
commutation relations unless $f(E)=1$).
However such a term can be eliminated with a
non-local redefinition of coordinates, $\xi_i =g(E)x_i$, with a
suitable function $g(E)$.
 In the second equation, again
$\delta_{ij}$ must appear because it is the only available tensor under
rotation. In order to recover the undeformed limit, we  further
require that $f(0)=1$ and that $a(E)$ is less singular than $E^{-2}$ as
$E\ra 0$. We neglect the possibility that the functions
$a,f$ depend also on other scalars like $x^2$ or ${\bf x\cdot p}$.

Of course, the form of the functions $a(E),f(E)$ is severely
restricted by the Jacobi identities.
Let us consider first  the Jacobi
identity $\left[ x_i,\left[ x_j,x_k\right]\right] +$ cyclic $=0$.
Using $\left[ x_i,E\right] =i\hbar f(E)p_i/E$, and
$\left[ x_i,a(E)\right] =i\hbar f(E) (p_i/E)da/dE$,
we get
\be
\frac{da}{dE}\, {\bf p}\cdot{\bf J}=0\, .
\ee
Since the Jacobi identity must be satisfied independently of the
particular representation
 of the  algebra, that is
independently of whether the condition ${\bf p}\cdot{\bf J}=0$ holds
or not, we conclude that $a(E)=$ const. With a  redefinition of
$\kappa$ we can set this constant to $\pm 1$.

The Jacobi identity $\left[ x_i,\left[ x_j,p_k\right]\right]
+$ cyclic $=0$ gives
\be
\frac{f(E)}{E}\frac{df}{dE}=\mp\frac{1}{\kappa^2}\, ,
\ee
where the upper (lower) sign correspond to the choice $a=+1 (-1)$.
Since $f(0)=1$, eq.~(5) gives $f(E)=(1\mp (E^2/\kappa^2))^{1/2}$.
All other Jacobi identities are automatically
 satisfied. Of course, to satisfy the Jacobi identities is in general
highly non-trivial, especially in a deformed algebra.
It is remarkable that in our case {\em i)} there exists a solution:
thus,  we can
$\kappa$-deform the Heisenberg algebra. {\em ii)} The solution is unique
(within our  assumptions and apart from the $\pm$ sign).
In the following we only consider the
lower sign\footnote{The solution with the upper sign is however quite
intriguing; since $f(E)$ is real by definition, the solution
is valid only for $E\le\kappa$ and  describes a system
which obeys standard quantum mechanics for $E\ll\kappa$ and
satisfies  $\left[ x,p\right] =0$ at $E=\kappa$.}
 and write the $\kappa$-deformed Heisenberg algebra as
\begin{eqnarray}
\left[ x_i ,x_j \right] &= & -\frac{\hbar^2}{\kappa^2}\,
i\epsilon_{ijk}J_k\label{xx}\\
\left[ x_i , p_j \right]   &= & i\hbar\delta_{ij}
(1+\frac{E^2}{\kappa^2})^{1/2}\, .\label{xp}
\end{eqnarray}
Actually, we already derived eqs.~(6,7)  in ref.~\cite{MM2}, following
a rather different route: we considered the $\kappa$-deformed
Poincar\'e algebra suggested in~\cite{LNR} and we discussed
how to generalize the  Newton-Wigner localization operator
to the $\kappa$-deformed case. We then obtained the algebra~(6,7),
with $x_i$ identified with the $\kappa$-deformed Newton-Wigner
operator and $p_i,J_i$ identified with the   generators of
spatial translations and rotations of the $\kappa$-deformed
Poincar\'e algebra. (To compare with
the results of ref.~\cite{MM2}
one must observe  that the present definition of $\kappa$ differs
by a factor of 2 from that used in~\cite{LNR,MM2} and that
in eq.~(\ref{xp}) $E^2$ is a shorthand for
${\bf p}^2+m^2$, while in the $\kappa$-Poincar\'e algebra the
dispersion relation is, with the present  definition of $\kappa$,
${\bf p}^2+m^2=\kappa^2\sinh^2 (E/\kappa )$.)

The derivation that we have presented in this Letter is instead
independent of whether we consider the usual or the
deformed Poincar\'e algebra.

We do not attempt to define a coproduct on the deformed
algebra given by eqs.~(6,7), since a system composed of two particles
localized at $x^{(1)}$ and $x^{(2)}$, respectively, cannot be
considered as a single localized system and therefore we cannot
require, in general, that it can be described by a coordinate
$X=X(x^{(1)},x^{(2)})$ which, togheter with the total momentum,
satisfies the same algebra as the constituents. Thus,
the  algebraic structure that we have presented  is not
a Hopf algebra, nor a quantum algebra.

{}From eq.~(\ref{xp}) we immediately
derive the generalized uncertainty principle
\be\label{gu}
 \Delta x_i\Delta p_j \ge  \frac{\hbar}{2} \delta_{ij}
\langle\left( 1+\frac{E^2}{\kappa^2}\right)^{1/2}
\rangle\, .
\ee
Expanding the square root in powers of
$(E/\kappa )^2$ and using
$\langle {\bf p}^2\rangle ={\bf p}^2+(\Delta p)^2$, where
$(\Delta p)^2=\langle ({\bf p}-\langle {\bf p}\rangle )^2\rangle$,
at first order we obtain
\be
 \Delta x_i\Delta p_j \ge  \frac{\hbar}{2} \delta_{ij}
\left( 1+\frac{E^2+(\Delta p)^2}{2\kappa^2}\right)
\, .
\ee
Thus, in the regime
$E\ll\kappa ,\Delta p\lsim\kappa$,
we  recover the string theory
result. The constant in  eq.~(1) is reproduced
if we identify $\hbar/\kappa$ with
the string length $\lambda_s$ times a numerical constant of order one.
Note that if
$\hbar/\kappa\sim\lambda_s$ (and, as usual, the string length is
larger than the Planck length),
then the condition $E\ll\kappa$ implies
the condition $2GE\ll\lambda_s$.

If we consider instead the regime
$\langle{\bf p}\rangle^2\sim (\Delta p)^2\gg\kappa^2$ we get
\be
\Delta x\ge {\rm const}\times \frac{\hbar}{\kappa}\, ,
\ee
while in the regime
$\langle {\bf p}\rangle^2\gg (\Delta p)^2$,
$\langle {\bf p}\rangle^2\gg\kappa^2$ we get
\be
\Delta x\ge {\rm const}\times \frac{\hbar}{\kappa}\,
\frac{|\langle{\bf p}\rangle |}{\Delta p}\gg \frac{\hbar}{\kappa}\, .
\ee

It is  important to observe  that  the
string theory result, eq.~(1), is reproduced in the appropriate
kinematical limit after expanding our closed expression at first order
in $\Delta p/\kappa$. However, only the full
expression, eq.~(8), has an underlying algebraic origin.

We now comment on the issue of Lorentz invariance.
When we discuss the
possible existence of a minimal observable length, we
actually  refer
to a spatial length. Such a concept, of course, is not Lorentz
invariant. We can always perform a boost and squeeze any 'minimal'
length as much as we like.  Then, if a minimal
spatial length actually exists, we must consider
 the very interesting possibility
that Lorentz invariance is not respected by
 quantum gravity at a scale on the order
of $\hbar/\kappa$.
In the formalism of the $\kappa$-deformed Poincar\'e
algebra of~\cite{LNR} this  is explicit, since the Lorentz algebra is
deformed by the parameter $\kappa$. In our
approach, it is implicit in the use of the three vectors $x_i,p_i$ in
place of the corresponding four-vectors. In this context it is
interesting to mention that a breaking of Lorentz invariance at the
string scale or at the Planck scale has been suggested very recently
by Susskind
in ref.~\cite{Sus}, where it is argued that Lorentz contraction of
particles must saturate as $E\sim M_{\rm PL}$, and a constant value
should be approached (cfr. our eqs.~(10,11)). It might be interesting
to observe that the considerations of ref.~\cite{Sus} stem from
Gedanken experiments with black holes, and a (different) Gedanken
experiment with black holes~\cite{MM1}
was also the starting point of our investigation.

Two final comments are in order. The first is that  we have a
scheme in which  Newton's
constant $G$ (or equivalently $\kappa$) enters  the theory at the
kinematical level (see also ref.~\cite{Tow} for early attempts in this
direction).  The minimal length also
emerges at a kinematical, rather than dynamical level.
This feature is rather satisfying. The second,
related comment, is that this formalism can acquire substance only
after a suitable dynamics has been added. This is particularly clear
if we repeat our analysis in $1+1$ dimensions. In this case only the
equation $\left[ x,p\right] =i\hbar f(E)$ survives. Of course, we can
always perform a non-local redefinition of the coordinate, defining
$\xi =x/f(E)$, so that $\left[\xi ,p\right] =i\hbar$. However, one
would pay this at the level of the action, which would become
non-local. The fact that a non-local particle Lagrangian can
generate a minimal observable length has been nicely shown
in~\cite{Kato}.

\vspace{5mm}

I thank M.~Mintchev, M.~Shifman and A.~Vainshtein for useful discussions.


\begin{thebibliography}{999}
\newcommand{\pl}{{ Phys. Lett.}\ }
\newcommand{\prl}{{Phys. Rev. Lett.}\ }
\newcommand{\pr}{{ Phys. Rev.}\ }
\newcommand{\np}{{ Nucl. Phys.}\ }
\newcommand{\jl}{{ JEPT Lett.}\ }
\newcommand{\js}{{ Sov. Phys. JEPT }\ }
\newcommand{\bb}{\bibitem}
\bb{GV} G. Veneziano, Europhys. Lett. 2 (1986) 199; Proc. of Texas
Superstring Workshop (1989).

D. Gross, Proc. of ICHEP, Munich (1988).
\bb{ACV1} D. Amati, M. Ciafaloni and G.~Veneziano,
\pl B197 (1987) 81, B216 (1989) 41;
Int.~J.~Mod. Phys. A3 (1988) 1615; \np B347 (1990) 530.
\bb{GM} D.J. Gross and P.F. Mende, \pl B197 (1987) 129; \np B303 (1988) 407.
\bb{KPP} K. Konishi, G. Paffuti and P.~Provero, \pl B234 (1990) 276;

R. Guida, K. Konishi and P.~Provero, Mod. Phys. Lett. A6 (1991) 1487.
\bb{Cia} M. Ciafaloni, preprint DFF 172/9/92.
\bb{MM1} M. Maggiore, \pl B304 (1993) 65.
\bb{Con} A. Connes, {\em Non-Commutative Differential Geometry},
IHES 62 (1986).
\bb{MM2} M. Maggiore, preprint IFUP-TH 19/93, hep-th/9305163, May 1993.
\bb{LNR} J. Lukierski, A. Nowicki and  H. Ruegg,
\pl  B293 (1992) 344.
\bb{Sus} L. Susskind, preprint SU-ITP-93-21, hep-th/9308139, Aug. 1993.
\bb{Tow} P.K. Townsend, \pr D15 (1977) 2795.
\bb{Kato} M.~Kato, \pl B245 (1990) 43.
\end{thebibliography}
\end{document}